# A new standard in high-field terahertz generation: the organic nonlinear optical crystal PNPA

Claire Rader[a], Zachary B. Zaccardi[a], (Enoch) Sin Hang Ho, Kylie G. Harrell, Paige K. Petersen, Megan F. Nielson, Harrison Stephan, Natalie K. Green, Daisy J.H. Ludlow, Matthew J. Lutz, Stacey J. Smith, David J. Michaelis*, and Jeremy A. Johnson*

Department of Chemistry and Biochemistry, Brigham Young University, Provo, UT 84602, USA

[a] Contributed equally to this work
* Electronic mail: dmichaelis@chem.byu.edu, jjohnson@chem.byu.edu

**Abstract:** We report the full characterization of a new organic nonlinear optical (NLO) crystal for intense THz generation: PNPA ((E)-4-((4-nitrobenzylidene)amino)-*N*-phenylaniline). We discuss crystal growth and structural characteristics. We present the wavelength dependence of THz generation, the thickness dependence of the THz spectrum for PNPA crystals, and measure the efficiency. PNPA enables intense THz generation that surpasses NLO crystals DAST and OH-1, which have been the standard in organic high-field THz generators for several years.

## 1. Introduction

The use of intense, broadband pulses of terahertz (THz) light as an ultrafast excitation source is enabling cutting-edge condensed-phase measurements. The relatively low-frequency THz light can resonantly excite collective modes, which efficiently drives motion far from equilibrium conditions. This efficient resonant excitation has enabled the direct measurement of anharmonic potential energy surfaces [1-3], the direct excitation and measurement of an electromagnon in multiferroic $TbMnO_3$ [4], and the transition to a metastable ferroelectric state in $SrTiO_3$ [5]. Experimental setups with the capability to generate a pair of intense THz pulses can also facilitate 2D THz spectroscopic measurements and have provided additional insights on a variety of fascinating systems [6-9].

Each of these experiments require high-field THz generation, and intense THz pulses were provided by organic nonlinear optical crystals like OH-1 [1, 2], DSTMS [2, 4, 8, 10], or DAST [11], or by inorganic $LiNbO_3$ in a tilted-pulse-front configuration [12]. DAST, the original organic crystal for high-field THz generation, was developed as a nonlinear optical crystal over 40 years ago [13]. Ten years ago, DAST was shown to be extremely efficient with THz generation, demonstrating peak field strengths in excess of 1 MV/cm [14]. Since that demonstration, several new THz generators have been developed, showing marginal improvement in THz output [10, 11, 15-18]. However, no new organic THz generator has significantly outperformed DAST.

Our laboratories recently disclosed a new data mining approach to discover and produce organic THz crystals [19]. We mined the Cambridge Structural Database (CSD) and combined the structural information with first-principles calculations to identify new organic materials with ideal molecular and solid-state properties for THz generation. Of these new materials, PNPA (((E)-4-((4-nitrobenzylidene)amino)-*N*-phenylaniline) was demonstrated to output the highest intensity THz generation, rivaling that generated by DAST and OH-1. Previous reports on the properties of PNPA had suggested its potential as a useful nonlinear optical crystal, but a full characterization of its THz generation capabilities has not been reported [19, 20]. In this report, we provide an in-depth analysis of the synthesis, growth, crystal characterization, and optical properties of PNPA and demonstrate that it enables THz generation that surpasses current state-of-the-art NLO crystals DAST and OH-1.

## 2. Experimental

Synthesis of PNPA: Equimolar amounts of 4-nitrobenzaldehyde (15.1 g, 0.1 mol) and 4-aminodiphenylamine (18.4 g, 0.1 mol) were added to a 500 mL flask. Next, 300 mL of ethanol and 3.0 g of anhydrous sodium sulfate were added. The reaction was stirred under an argon atmosphere at 80 °C for 1 h. The ethanol was removed by vacuum evaporation (rotovap) and 300 mL of ethanol was again added to the reaction. The mixture was again stirred for 1 h at 80 °C, followed by removal of solvent. The solid product was then dissolved in 1 L of $CH_2Cl_2$ and the solution was filtered through a silica gel column using $CH_2Cl_2$ as eluent to provide the desired product as a solid (31.7 g, 100% yield). M.P. = 176.5 C. $^1$H NMR (500 MHz, CDCl$_3$), δ 8.61 (s, 1H), 8.33-8.31 (d, $J$ = 8.8, 2H), 8.07-8.05 (d, $J$ = 8.8, 2H), 7.33-7.30 (m, 4H), 7.12-7.11 (m, 4H), 7.00-6.98 (t, $J$ = 7.4, 1H), 5.86 (s, 1H).

Large Crystal Growth: Previous studies have reported that large PNPA crystals can be grown from chloroform and methanol mixtures via slow evaporation [20]. Our additional studies demonstrate that crystal growth via slow evaporation from $CH_2Cl_2$ provides the best quality crystals with the largest sizes. In a typical slow evaporation experiment, 3.3 g of powdered (pure) PNPA was dissolved in 100 mL of $CH_2Cl_2$ with heating and stirring until all solids dissolved. The resulting solution was allowed to slowly evaporate at ~1.5 mL per day at room temperature over the course of 2 weeks. During this time, red, prismatic seed crystals began to form. The resulting seed crystals were extracted after the 2 weeks and resubjected to a solution concentrated at 3.1 g of PNPA per 100 mL of $CH_2Cl_2$ at the same rate and for the same amount of time. Resulting crystals recovered from solution ranged in size from 5 mm × 10 mm to 10 mm × 10 mm with varying thicknesses due to the prismatic shape of the crystals. Under the same conditions, some flat crystals were grown with uniform thickness. High quality crystal windows of varying thickness were then cleaved from the prisms and polished to desired thicknesses. Figure 1 shows examples of (a) prismatic as-grown PNPA crystals and (b) a cleaved and polished crystal.

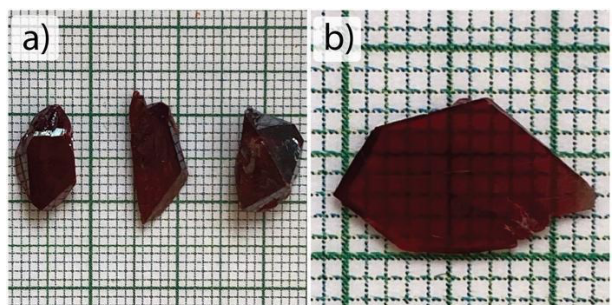

Figure 1. a). As grown prismatic PNPA crystals from slow evaporation in $CH_2Cl_2$. b) Cleaved and polished PNPA crystal window used for THz generation.

Structure confirmation and face determination: Single-crystal X-ray diffraction was used to confirm that our PNPA grown under the above method maintained the same crystal structure as the reported structure in the database (CCDC ID: 991743) [19]. We also identified the main crystal face for irradiation (010) using powder X-ray diffraction (See section 2 in supporting information for experimental details).

Terahertz generation measurements: Measurements were performed using ultrafast pump pulses with a broad range of NIR wavelengths. To generate these wavelengths, the 800-nm light generated from a Ti:sapphire regenerative amplifier is sent into an optical parametric amplifier which outputs light across the whole the range from 1250 to 2150 nm (1250-1600 nm signal, 1600-2150 nm idler). The signal or idler pump pulses (0.38 cm $1/e^2$ radius, ~100 fs pulse duration) are directed to the THz generation crystal and excess pump light is removed using a 2 mm thick HDPE filter. The transmitted THz is expanded using a 1:5 telescope consisting of 1-inch diameter, 1-inch reflected focal length and 3-inch diameter, 5-inch reflected focal length off-axis gold-coated parabolic mirrors. A final 3-inch diameter, 2-inch reflected focal length off-axis parabolic mirror focuses the generated THz down to a 260 μm beam radius. We record THz electric-field traces at the focus using a 100-μm (110) electro-optic GaP crystal bonded to a 1-mm (001) GaP crystal and probe with 100 fs 800-nm light pulses in a standard electro-optic configuration [21].

## 3. Results and Discussion
**Molecular Property and X-ray Crystal Structure**
Simple models show that three of the key factors that govern the THz generation efficiency in a NLO crystal include the molecular hyperpolarizability, molecular alignment in the crystalline state, and the irradiated crystal face. Through our initial data mining effort, the hyperpolarizability of PNPA molecules was calculated to be $316 \times 10^{-30}$ esu [19], which is significantly larger than the values for OH-1 ($89.6 \times 10^{-30}$ esu) and DAST ($201 \times 10^{-30}$ esu).

In order to quantify the molecular alignment, the crystal packing order parameter ($\cos^3(\theta_p)$) was calculated, where $\theta_p$ is defined as the angle between the molecular hyperpolarizability vector and the polar axis of the unit cell [22]. An order parameter of 1 indicates an ideal head-to-tail molecular alignment and a value of 0 indicates centrosymmetric packing without NLO activity. The value of $\theta_p$ for PNPA is 0.66°, giving an order parameter is 0.9998 (see Figure 2). Compared to the order parameter values of OH-1 (0.69) and DAST (0.83), PNPA has a nearly ideal molecular alignment for THz generation.

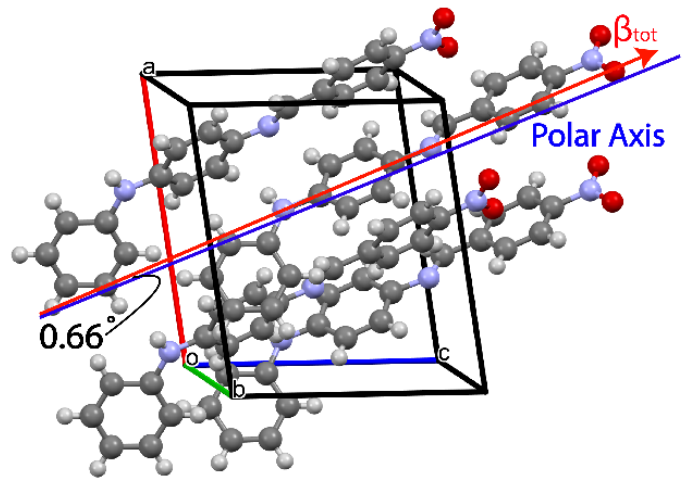

Figure 2. (a) Crystal packing of PNPA and the angle between polar axis (blue) and hyperpolarizability vector (red). The indicated view is along the crystal face (010) of our PNPA crystal, which is the optimal face for THz generation.

**THz Generation and Optical Characterization**
The near ideal non-centrosymmetric packing and high hyperpolarizability of PNPA make it a strong candidate for high intensity THz generation, which we predicted to surpass both DAST and OH-1. Here we provide a comprehensive description of the THz generation capabilities of PNPA and we directly compare PNPA to current standard THz generators DAST and OH-1.

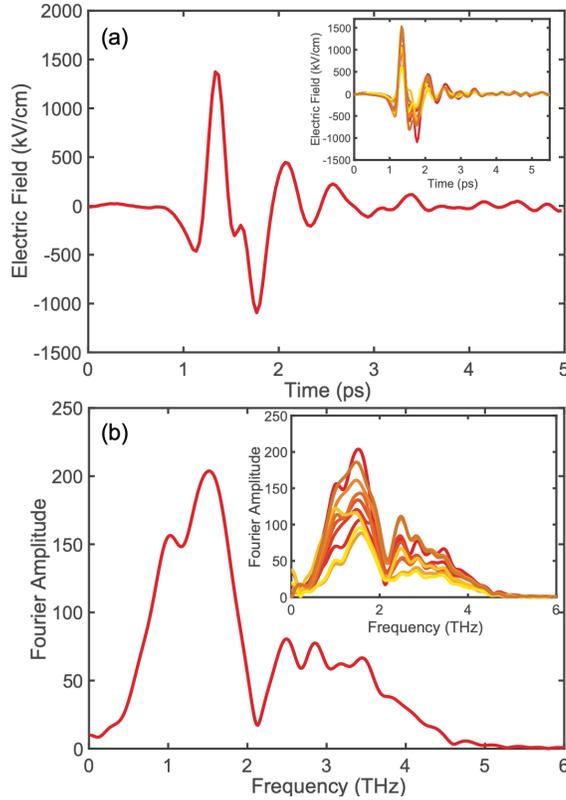

Figure 3. (a) The time-dependent electric field of PNPA, with an inset that shows time traces of ten PNPA crystals. (b) The spectrum of the PNPA electric field from (a), with an inset showing the spectra of ten PNPA crystals.

PNPA generates a strong, single-cycle pulse with the capability to reach peak-to-peak electric-field strengths >2 MV/cm, as shown in Figure 3a. Figure 3b shows the associated spectrum with an absorption dip at 2 THz and frequencies extending out to 5 THz. We also tested multiple crystals to ensure that crystal growth was replicable and led to consistent THz output. The insets of Figure 3 show THz traces and spectra of ten PNPA crystals, where small variabilities arise from differences in crystal thickness and quality. In all of these measurements, PNPA was irradiated with 1450 nm pump wavelength and a pulse energy of ~0.6 mJ.

In Figure 4, we show THz generation measured from five PNPA crystals of increasing thickness, ranging from 501 μm up to 1.24 mm. As shown in Figure 4a, the THz amplitude below 2 THz increases as the crystal thickness increases, while above 2 THz, we see similar spectral amplitudes. These trends indicate a long coherence length for frequencies <2 THz and a generation length <500 μm for frequencies >2 THz. The inset of Figure 4a shows the normalized Fourier transform for each crystal. Figure 4b shows the electric-field measurements for the spectra in 4a. The inset to 4b shows the peak-to-peak electric field strength versus thickness.

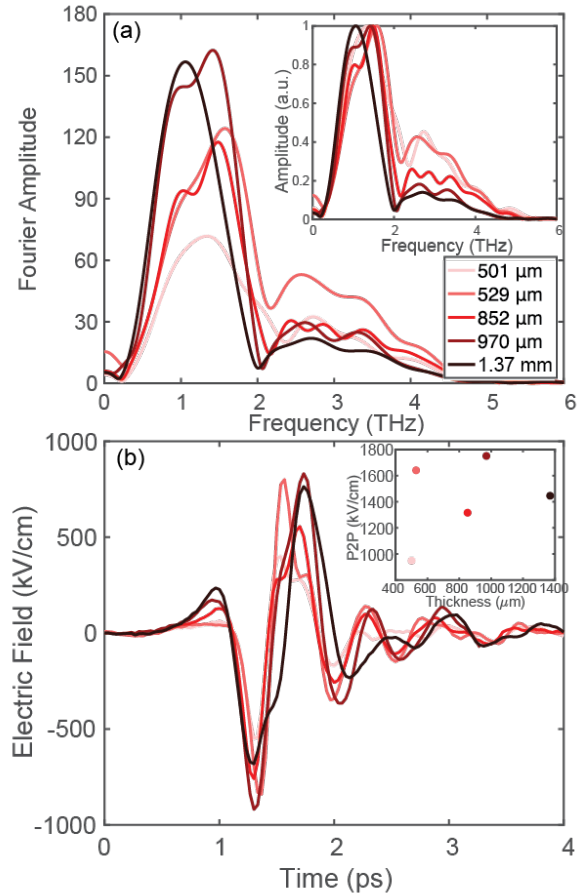

Figure 4. Thickness dependent data for PNPA. (a) Fourier transform of each crystal thickness, with normalized spectra in inset. (b) Electric-field traces, with inset showing peak-to-peak field strengths versus crystal thickness.

Wavelength dependent measurements highlight another useful aspect of PNPA. While most organic generation crystals perform best in a select range of pump wavelengths, PNPA can be used across a broad range of wavelengths, as shown in Figure 5. Figure 5a shows the THz spectra recorded when generating with a 529 μm thick crystal and varying the pump wavelength from 1250 nm to 2150 nm in 100 nm steps. The pump pulse energy was held constant at ~0.26 mJ. Each spectrum exhibits a peak in spectral amplitude at ~1.5 THz, followed by an absorption dip just above 2 THz. There is significant amplitude up to higher frequencies, however the response is limited by the pump pulse duration. Figure 5b shows peak-to-peak electric-field versus pump wavelength. We see that from 1250 nm to 1800 nm there is a fairly consistent peak-to-peak THz field strength, with an approximately 10% increase leading to a maximum output at 1565 nm. From 1800 nm to 2150 nm, the field strength monotonically decreases to roughly half the maximum value. The inset to Figure 5b shows the percent transmission of light through PNPA as a function of wavelength from 500 nm to 2000 nm. The absorption cutoff at 611 nm is marked with a yellow dot. The dip in the peak-to-peak electric field at 1500 nm can be explained by the absorption of the pump light at 1500 nm.

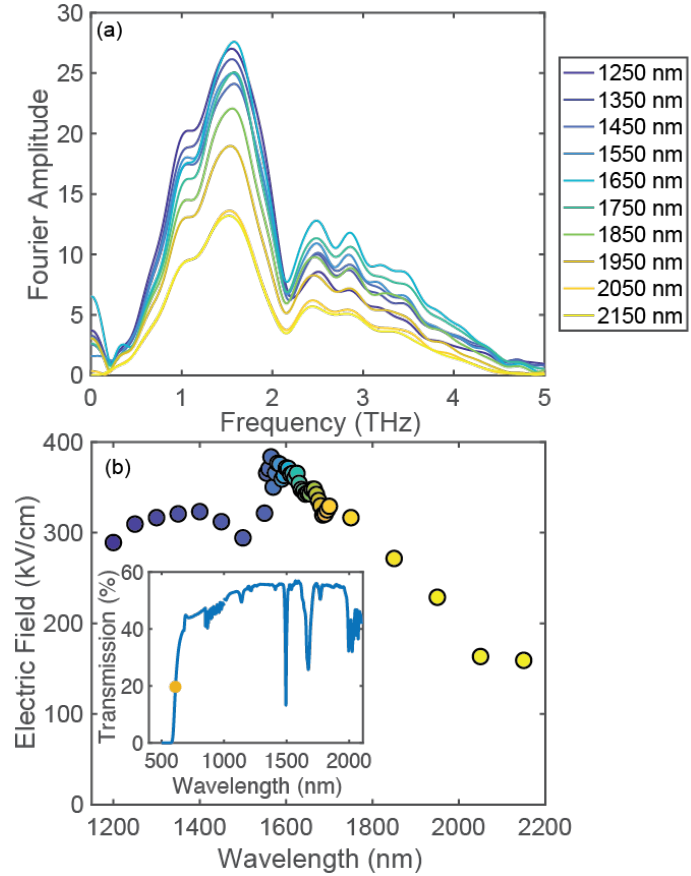

Figure 5. (a) The Fourier spectrum of wavelength dependent THz traces pumped from 1250 nm to 2150 nm. (b) Peak-to-peak electric-field strength versus pump wavelength, showing that PNPA performs best in the range 1250 nm to 1800 nm. Inset shows percent of pump light transmitted through PNPA as a function of wavelength.

We next characterized how PNPA generates THz as a function of pump fluence. To extend the range of accessible pump fluences, we used a telescope to reduce the pump spot size to a $1/e^2$ radius of 0.19 cm. With this spot size and a pulse energy of ~0.63 mJ at 1450 nm, the highest fluence reached was ~5.5 mJ/cm$^2$. Visual inspection and repeat use of the crystal showed no damage to the crystal. To more carefully characterize the damage threshold, we would need to measure with higher fluences. In Figure 6, we show the pump-fluence dependence of PNPA, directly compared to DAST and OH-1 with the same experiment conditions.

Saturation effects start to appear in PNPA at ~2 mJ/cm$^2$, as suggested by the dotted red line. We also characterized the generation efficiency of all three crystals, as shown in the inset. These measurements demonstrate that PNPA generates significantly more THz than DAST and OH-1. We see peak-to-peak field strengths approaching 3 MV/cm, as well as a higher efficiency.

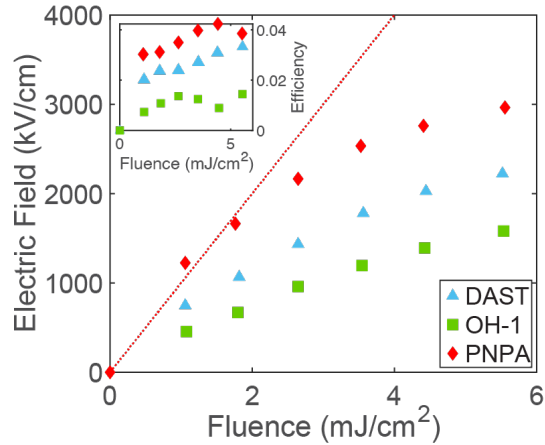

Figure 6. Fluence dependence of PNPA (red diamond), DAST (blue triangles) and OH-1 (green square). The dotted red line is the fitting for the PNPA measurement. Efficiency measurements are shown in the inset using the same color scheme.

Figure 6 shows a clear advantage of using PNPA over DAST and OH-1, however using a single value to characterize broadband THz generation is inadequate. To further elaborate, Figure 7 shows the THz traces (Fig. 7a) and spectra (Fig. 7b) of the same crystals recorded at the maximum fluence conditions in Figure 6. PNPA exceeds DAST and OH-1 in both the maximum field strengths achieved and in spectral breadth. DAST and OH-1 respectively reach peak-to-peak electric field strengths of 2.2 MV/cm and 1.5 MV/cm, while PNPA reaches 2.9 MV/cm. The combination of an exceedingly high field strength and large spectral amplitudes across a broad range of frequencies confirms that PNPA should be used as a new standard in high-field THz generation.

We also note that PNPA is more amenable to growing large crystals and polishing smooth surfaces than DAST and OH-1, which can crack more easily during the polishing process. This advantage makes PNPA not only an extremely efficient THz generator, but a more synthetically accessible organic NLO crystal as well.

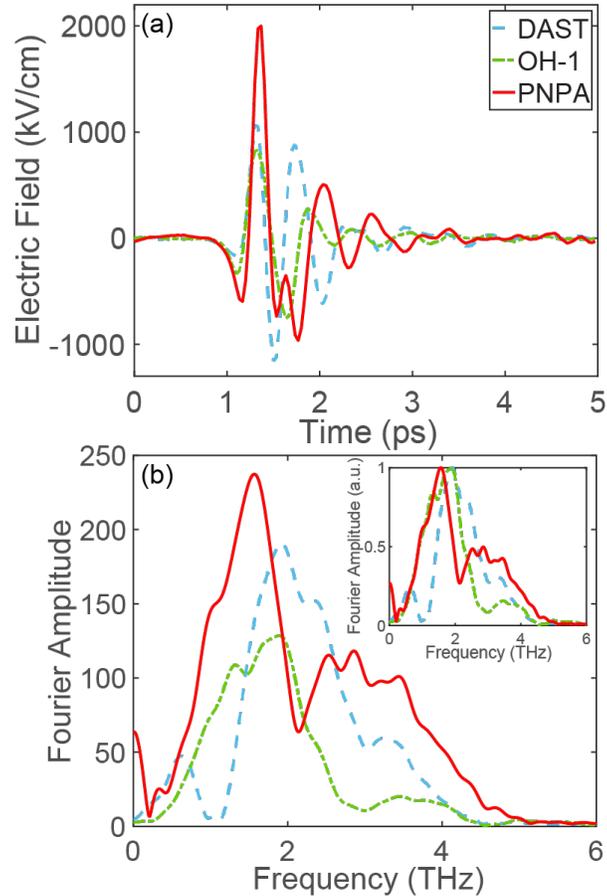

Figure 7. a) The time trace and (b) spectra of PNPA (solid red) compared with that of DAST (dashed blue) and OH-1 (dash-dotted green). Normalized spectra of the three crystals are in the inset of (b).

### 4. Conclusion

In summary, we demonstrate crystallization methods for PNPA that enable access to high quality single crystals for THz generation applications. When comparing PNPA to crystals that have been the standards in high-field THz generation for several years (DAST and OH-1), we find that PNPA outperforms these standards, granting access to much higher electric field strengths. We also present thickness and wavelength dependent data, showing that PNPA performs well over a broad range of wavelengths. These results demonstrate the utility of PNPA for THz generation compared to other commercial organic NLO crystals and sets a new standard for high-field THz generation.

### 5. References

1.　B. S. Dastrup, J. R. Hall, and J. A. Johnson, Experimental determination of the interatomic potential in LiNbO3 via ultrafast lattice control, Applied Physics Letters **110**, 162901 (2017), https://aip.scitation.org/doi/abs/10.1063/1.4980112.
2.　B. E. Knighton, R. T. Hardy, C. L. Johnson, L. M. Rawlings, J. T. Woolley, C. Calderon, A. Urrea, and J. A. Johnson, Terahertz waveform considerations for nonlinearly driving lattice vibrations, Journal of Applied Physics **125**, 144101 (2019), https://aip.scitation.org/doi/abs/10.1063/1.5052638.
3. M. Kozina, M. Fechner, P. Marsik, T. van Driel, J. M. Glownia, C. Bernhard, M. Radovic, D. Zhu, S. Bonetti, U. Staub, and M. C. Hoffmann, Terahertz-driven phonon upconversion in SrTiO3, Nature Physics **15**, 387-392 (2019), https://doi.org/10.1038/s41567-018-0408-1.
4. T. Kubacka, J. A. Johnson, M. C. Hoffmann, C. Vicario, S. d. Jong, P. Beaud, S. Grübel, S.-W. Huang, L. Huber, L. Patthey, Y.-D. Chuang, J. J. Turner, G. L. Dakovski, W.-S. Lee, M. P. Minitti, W. Schlotter, R. G. Moore, C. P. Hauri, S. M. Koohpayeh, V. Scagnoli, G. Ingold, S. L. Johnson, and U. Staub, Large-Amplitude Spin Dynamics Driven by a THz Pulse in Resonance with an Electromagnon, Science **343**, 1333-1336 (2014), https://www.science.org/doi/abs/10.1126/science.1242862
5. X. Li, T. Qiu, J. Zhang, E. Baldini, J. Lu, A. M. Rappe, and K. A. Nelson, Terahertz field–induced ferroelectricity in quantum paraelectric $SrTiO_3$, Science **364**, 1079-1082 (2019), https://www.science.org/doi/abs/10.1126/science.aaw4913
6. I. A. Finneran, R. Welsch, M. A. Allodi, T. F. Miller, and G. A. Blake, Coherent two-dimensional terahertz-terahertz-Raman spectroscopy, Proceedings of the National Academy of Sciences **113**, 6857-6861 (2016), https://www.pnas.org/doi/abs/10.1073/pnas.1605631113
7. I. A. Finneran, R. Welsch, M. A. Allodi, T. F. Miller, and G. A. Blake, 2D THz-THz-Raman Photon-Echo Spectroscopy of Molecular Vibrations in Liquid Bromoform, The Journal of Physical Chemistry Letters **8**, 4640-4644 (2017), https://doi.org/10.1021/acs.jpclett.7b02106.
8. C. L. Johnson, B. E. Knighton, and J. A. Johnson, Distinguishing Nonlinear Terahertz Excitation Pathways with Two-Dimensional Spectroscopy, Physical Review Letters **122**, 073901 (2019), https://link.aps.org/doi/10.1103/PhysRevLett.122.073901.
9. J. Lu, X. Li, H. Y. Hwang, B. K. Ofori-Okai, T. Kurihara, T. Suemoto, and K. A. Nelson, Coherent Two-Dimensional Terahertz Magnetic Resonance Spectroscopy of Collective Spin Waves, Physical Review Letters **118**, 207204 (2017), https://link.aps.org/doi/10.1103/PhysRevLett.118.207204.
10. S. Houver, L. Huber, M. Savoini, E. Abreu, and S. L. Johnson, 2D THz spectroscopic investigation of ballistic conduction-band electron dynamics in InSb, Optics Express **27**, 10854-10865 (2019), http://opg.optica.org/oe/abstract.cfm?URI=oe-27-8-10854.
11. G. Mead, H.-W. Lin, I.-B. Magdău, T. F. Miller, and G. A. Blake, Sum-Frequency Signals in 2D-Terahertz-Terahertz-Raman Spectroscopy, The Journal of Physical Chemistry B **124**, 8904-8908 (2020), https://doi.org/10.1021/acs.jpcb.0c07935.
12. H. Hirori, A. Doi, F. Blanchard, and K. Tanaka , "Single-cycle terahertz pulses with amplitudes exceeding 1 MV/cm generated by optical rectification in LiNbO3 $LiNbO_3$", Appl. Phys. Lett. 98, 091106 (2011) https://doi.org/10.1063/1.3560062
13. J. Perry, S. Marder, K. Perry, E. Sleva, C. Yakymyshyn, K. Stewart, and E. Boden, *Organic salts with large electro-optic coefficients* (SPIE, 1991).
14. B. Monoszlai, C. Vicario, M. Jazbinsek, and C. P. Hauri, High-energy terahertz pulses from organic crystals: DAST and DSTMS pumped at Ti:sapphire wavelength, Optics Letters **38**, 5106-5109 (2013), http://opg.optica.org/ol/abstract.cfm?URI=ol-38-23-5106.
15. G. A. Valdivia-Berroeta, L. K. Heki, E. A. McMurray, L. A. Foote, S. H. Nazari, L. Y. Serafin, S. J. Smith, D. J. Michaelis, and J. A. Johnson, Alkynyl Pyridinium Crystals for Terahertz Generation, Advanced Optical Materials **6**, 1800383 (2018), https://onlinelibrary.wiley.com/doi/abs/10.1002/adom.201800383.

16. S.-H. Lee, J. Lu, S.-J. Lee, J.-H. Han, C.-U. Jeong, S.-C. Lee, X. Li, M. Jazbinšek, W. Yoon, H. Yun, B. J. Kang, F. Rotermund, K. A. Nelson, and O.-P. Kwon, Benzothiazolium Single Crystals: A New Class of Nonlinear Optical Crystals with Efficient THz Wave Generation, Advanced Materials **29**, 1701748 (2017), https://onlinelibrary.wiley.com/doi/abs/10.1002/adma.201701748.
17. B. J. Kang, I. H. Baek, S.-H. Lee, W. T. Kim, S.-J. Lee, Y. U. Jeong, O. P. Kwon, and F. Rotermund, Highly nonlinear organic crystal OHQ-T for efficient ultra-broadband terahertz wave generation beyond 10 THz, Optics Express **24**, 11054-11061 (2016), http://opg.optica.org/oe/abstract.cfm?URI=oe-24-10-11054.
18. S.-H. Lee, S.-J. Lee, M. Jazbinsek, B. J. Kang, F. Rotermund, and O. P. Kwon, Electro-optic crystals grown in confined geometry with optimal crystal characteristics for THz photonic applications, CrystEngComm **18**, 7311-7318 (2016), http://dx.doi.org/10.1039/C6CE00958A.
19. G. A. Valdivia-Berroeta, Z. B. Zaccardi, S. K. F. Pettit, S.-H. Ho, B. W. Palmer, M. J. Lutz, C. Rader, B. P. Hunter, N. K. Green, C. Barlow, C. Z. Wayment, D. J. Ludlow, P. Petersen, S. J. Smith, D. J. Michaelis, and J. A. Johnson, Data Mining for Terahertz Generation Crystals, Advanced Materials **34**, 2107900 (2022), https://onlinelibrary.wiley.com/doi/abs/10.1002/adma.202107900.
20. N. Kooliyankal, R. Sreedharan, S. Ravi, R. K. Ravindran, and M. K. T. Krishnankutty, Experimental and computational studies on optical properties of a promising N-benzylideneaniline derivative for non-linear optical applications, Zeitschrift für Naturforschung A **75**, 557-573 (2020), https://doi.org/10.1515/zna-2020-0047.
21. Q. Wu, and X.-C. Zhang, 7 terahertz broadband GaP electro-optic sensor, Applied Physics Letters **70**, 1784-1786 (1997), https://aip.scitation.org/doi/abs/10.1063/1.118691.
22. J. Zyss, and J. L. Oudar, Relations between microscopic and macroscopic lowest-order optical nonlinearities of molecular crystals with one- or two-dimensional units, Physical Review A **26**, 2028-2048 (1982), https://link.aps.org/doi/10.1103/PhysRevA.26.2028.